\documentclass[12pt]{article}
\usepackage{graphicx,epstopdf,amssymb,amsfonts,amsmath,amsthm,array,
mathrsfs,amscd,flafter,mathtools,cite}%,wick}%flafter force les figures à apparaître après l'endroit où elles sont définies
\usepackage{hyperref}% hyperliens dans le fichier .pdf
\usepackage{todonotes} %remarques dans la marge avec la commande \todo
\newcommand{\pbs}[1]{\let\temp=\\#1\let\\=\temp}
\numberwithin{equation}{section}
\def\be{\begin{equation}}\def\ee{\end{equation}}
\def\cvp{\raise 2pt\hbox{,}}

\def\tr{\mathop{\rm tr}\nolimits}

\def\d{{\rm d}}
\def\nn{{\cal N}}

\def\Seff{S_{\text{eff}}}

\def\SO{\text{SO}}

\def\gs{g_{\text{s}}}
\def\ls{\ell_{\text{s}}}
\def\1{\mathbb{I}}

\def\UK{{\text{U($K$)}}}
\def\UN{\text{U($N$)}}

\def\truK{\mathop{\text{tr}_{\text{U}(K)}}\nolimits}

\def\del{\partial}
\def\z{\zeta}

\def\e{\epsilon}
\begin{document}
{\pagestyle{empty}
\parskip 0in
\
\vfill
\begin{center}
{\LARGE  Emergent D5-Brane Background from D-Strings}\\
\vspace{0.4in}
Frank F{\scshape errari}
and Antonin R{\scshape ovai}
\\
\medskip
{\it Service de Physique Th\'eorique et Math\'ematique\\
Universit\'e Libre de Bruxelles and International Solvay Institutes\\
Campus de la Plaine, CP 231, B-1050 Bruxelles, Belgique}
\smallskip
{\tt frank.ferrari@ulb.ac.be, antonin.rovai@ulb.ac.be}
\medskip
\end{center}
\vfill\noindent

We solve the worldsheet theory describing the near-horizon dynamics of a D-string in the presence of a very large number $N$ of D5-branes. The model is pre-geometric in the sense that the near-horizon worldsheet Lagrangian does not have dynamical fields associated with the dimensions transverse to the D5-branes. The solution at large $N$ is shown to be given by a classical action for the D-string moving in a curved ten-dimensional spacetime. The four dimensions transverse to the D5-branes emerge from the quantum loops of the original strongly coupled quantum worldsheet field theory. By comparing with the Dirac-Born-Infeld plus Chern-Simons action for a D-string in a general type IIB background, we identify the string-frame metric, dilaton and Ramond-Ramond three-form field-strength and find a match with the near-horizon geometry of a stack of D5-branes. 

\vfill
\medskip
\begin{flushleft}
\today
\end{flushleft}
\newpage\pagestyle{plain}
\baselineskip 16pt
\setcounter{footnote}{0}}
\section{Introduction}
\label{introduction}

The AdS/CFT correspondence \cite{Maldacena:1997re,Gubser:1998bc,Witten:1998qj} and its generalizations (see e.g.\ \cite{Aharony:1999ti,D'Hoker:2002aw} for reviews and references) offer a framework in which, at least in principle, geometry and quantum gravity can be studied from ordinary quantum field gauge theories. In this framework, a $d$-dimensional field theory is postulated to be dual, in the large $N$ limit, to a classical gravitational theory defined on a higher $D>d$ dimensional curved spacetime. The $D-d$ additional dimensions are emerging from strongly coupled quantum physics in the gauge theory. The study of the emerging geometry from the pure field theory point of view is unfortunately rather difficult in general. It requires to compute observables in the strong coupling regime of the field theory and to extract from them informations about the dual geometry.

Recently, it was proposed in \cite{Ferrari:2012nw} to focus on the field theory models describing the near-horizon worldvolume dynamics of a fixed number of D-branes, called the probe branes, in the presence of a very large number $N$ of other D-branes, called the background branes.
These models are pre-geometric, because they do not contain dynamical fields associated with the dimensions transverse to the background branes. According to the AdS/CFT lore, in the large $N$ limit, they should be equivalent to classical worldvolume theories describing the motion of the probe D-branes in the ten dimensional supergravity solution sourced by the background branes.
This non-trivial supergravity solution should thus emerge, alongside the transverse dimensions of space on which it lives, from the large $N$ solution of the probe branes worldvolume model. Much more details on these ideas can be found in \cite{Ferrari:2012nw,Ferrari:2013pq,Ferrari:2013pi}.

The power of this approach is that it is easier to extract some strong coupling information from the large $N$ probe brane models than from the large $N$ background brane theories \cite{Ferrari:2012nw}: the latter are complicated matrix models, whereas the former are vector-like models. It is then possible to explicitly \emph{derive} the emergence of space from a microscopic calculation in the strongly coupled vector-like worldvolume theory. In all the examples that have been studied so far \cite{Ferrari:2012nw,Ferrari:2013pq,Ferrari:2013hg}, the solution at large $N$ can be described by a classical action matching the non-abelian D-brane action for the probe branes in the correct ten-dimensional curved background. The scalar fields associated with the emerging classical dimensions of space are composite variables in the original microscopic description whose quantum fluctuations are suppressed in the large $N$ limit by the quantum loops of the vector variables.

The aim of the present work is to apply the above ideas to derive the emergent near-horizon geometry of a large number of D5-branes in type IIB string theory by studying the corresponding probe D-string worldsheet model. We start in Section \ref{action} by presenting in details the relevant D-string microscopic pre-geometric worldsheet theory. We solve the model at large $N$ in Section \ref{solution} and find that the solution is expressed in terms of a classical action which contains the right dynamical fields to describe the motion of the D-strings in a ten-dimensional background. In Section \ref{comparison}, we compare the expansion of this action around a flat worldsheet with the corresponding terms derived from the well-known Dirac-Born-Infeld plus Chern-Simons D-string action in arbitrary supergravity background. This allows us to identify the string-frame metric, dilaton and Ramond-Ramond three-form field-strength from a purely field theoretic calculation. The result matches perfectly the near-horizon supergravity solution sourced by the background D5-branes.

\section{The D-string microscopic action}
\label{action}

In principle, we need the action describing the dynamics of the open string modes of a system composed of a fixed number $K$ of D-strings and a large number $N$ of  D5-branes, in an appropriate decoupling, near-horizon limit \cite{Itzhaki:1998dd}. All the branes are chosen to be parallel to each other, and we work in the Euclidean. In particular, the model preserves eight supercharges and there is an 
$\text{SO}(2)\times\text{SO}(4)
\times\text{SO}(4)'$ global symmetry group corresponding to rotations in spacetime preserving the brane configuration: $\text{SO}(2)$ is associated with rotations on the D-string worldsheet, $\text{SO}(4)'$ with rotations on the D5-brane worldvolume transverse to the D-strings and $\text{SO}(4)$ with rotations transverse to both the D-strings and the D5-branes.

This action could be studied by evaluating appropriate low-energy limits of open string disk diagrams with various boundary conditions. This is a very tedious and complicated procedure, which was performed in the case of the D(-1)/D3 system in \cite{Green:2000ke,Billo:2002hm}. In our case, the result is a sum of a worldsheet action for the D-strings and a worldvolume action for the D5-branes, with couplings between the D-string and D5-brane degrees of freedom. However, the na\"\i ve action obtained in this way 
for the degrees of freedom living on the D5-branes would not be renormalizable and thus could be used only in the infrared. In other words, the full description of the D5-brane is not field theoretic. Fortunately, an explicit description of the D5-brane degrees of freedom and their couplings to the D-strings will not be required for our purposes. Indeed, due to supersymmetry, these couplings are not expected to contribute to the terms in the effective action on which we shall focus. A similar non-renormalization theorem was discussed in \cite{Diaconescu:1997ut} in the case of D-particles. We refer to \cite{Ferrari:2012nw} for a detailed discussion of these issues in the case of the D(-1)/D3 system and to \cite{frankonematrix} for a general discussion in non-supersymmetric contexts.

We thus focus on the D-string worldsheet Lagrangian, without referring any longer to possible couplings to the D5-brane fields. The form of the Lagrangian is strongly contrained by $(4,4)$ supersymmetry and global symmetries.
The simplest way to derive it is to proceed in two steps. First, we perform the dimensional reduction of the $\UK$ $\nn=1$ gauge theory in six dimensions down to two dimensions. The theory must contain one hypermultiplet in the adjoint, corresponding to D1/D1 string degrees of freedom, and $N$ hypermultiplets in the fundamental corresponding to the D1/D5 strings. Second, a suitable scaling limit must be implemented on this action, which corresponds to the correct low energy limit associated with the standard Maldacena near-horizon limit \cite{Maldacena:1997re,Itzhaki:1998dd,WittenCFTHiggs}. We present these two steps successively in the next two subsections.

\subsection{Dimensional reduction from six to two dimensions}
\label{sixtotwo}

The six dimensional $\text{U}(K)$ gauge theory has a spacetime symmetry group $\text{SO}(6)$, R-symmetry group $\text{SU}(2)_{+}$, internal symmetry group $\text{SU}(2)_{-}$ and flavor symmetry group $\text{SU}(N)$. The field content is as follows. The vector multiplet is composed of the gauge potential $A_{r}$, $1\leq r\leq 6$ and the gluino $\Lambda_{\alpha}$, which is a doublet of $\text{SU}(2)_{+}$. The adjoint hypermultiplet contains the adjoint scalars $a_{\mu}$ in the vector of $\text{SO}(4)\sim \text{SU}(2)_{+}\times \text{SU}(2)_{-}$ and a fermionic doublet $\bar\Lambda^{\dot\alpha}$ of $\text{SU}(2)_{-}$. The fundamental hypermultiplets contain $\text{SU}(2)_{+}$ scalar doublets $(q^{\alpha}_{f},\tilde q^{\alpha f})$ and Weyl fermions $(\chi_{f},\tilde\chi^{f})$ in the fundamental and antifundamental of $\text{SU}(N)$. The chirality of the Weyl spinors in the vector multiplet on the one hand and in the hypermultiplets on the other hand must be opposite.

After the dimensional reduction down to two dimensions, the six-dimensional spacetime symmetry group $\text{SO}(6)$ yields the two-dimensional spacetime symmetry group $\text{SO}(2)$ and a new global $\text{SO}(4)'\sim\text{SU}(2)'_{+}\times\text{SU}(2)'_{-}$. The fields of the six-dimensional theory  are then reorganized into representations of $\text{SO}(2)$ and $\text{SO}(4)'$, while their transformation laws under $\UK$, $\UN$, $\text{SU}(2)_{+}$ and $\text{SU}(2)_{-}$ are left unchanged.

More precisely, the six dimensional vector multiplet yields the two dimensional gauge field $A_{I}$, four scalars $\phi_{m}$ transforming in the vector representation of $\text{SO}(4)'$ and spinors $\Lambda_{\alpha \z}$, $\Lambda_{\alpha\dot\z}$ transforming in the representations $(1/2,0)_{1/2}$ and $(0,1/2)_{-1/2}$ of $\text{SO}(4)'\times\text{SO(2)}$. The adjoint hypermultiplet yields four scalars $a_{\mu}$ and spinors $\bar\Lambda_{\dot\alpha \z}$, $\bar\Lambda_{\dot\alpha\dot\z}$ transforming in the representations $(1/2,0)_{-1/2}$ and $(0,1/2)_{1/2}$ of $\text{SO}(4)'\times\text{SO(2)}$. Finally, the fundamental hypermultiplets yield the scalars $q^{\alpha}_{f}$ and $\tilde q^{\alpha f}$ together with fermions \smash{$(\chi_{\z f},\tilde\chi^{\phantom{\z}f}_\z)$} in the $(1/2,0)_{-1/2}$ and \smash{$(\chi_{\dot\z f},\tilde\chi_{\dot\z}^{\ f})$} in the $(0,1/2)_{1/2}$ representations of $\text{SO}(4)'\times\text{SO(2)}$ respectively. 
Let us note that the model is also invariant under worldsheet parity transformations which act by exchanging the $\text{SU}(2)'_{+}$ and 
$\text{SU}(2)'_{-}$ factors of $\text{SO}(4)'$. All these symmetry properties are summarized in Table \ref{recapsym} in the Appendix.

The $(4,4)$ supersymmetric Lagrangian resulting from the dimensional reduction can then be written as
\begin{multline}\label{d1lag}
	\tilde L = \frac{\ls^{2}}{2\gs} \truK \biggl( \frac{2}{\ls^{4}}\1_K+ \frac 12 F_{IJ}F_{IJ} + \nabla_{\! I} \phi_m\nabla_{\! I} \phi_m -\frac 12 \bigr[\phi_m,\phi_n\bigl]\bigl[\phi_m,\phi_n\bigr] + \nabla_{\! I} a_\mu \nabla_{\! I} a_\nu  
	\\
	- \bigl[\phi_m,a_\mu \bigr]\bigl[\phi_m,a_\mu \bigr] + 2i \bigl[a_\mu,a_\nu \bigr]D_{\mu\nu} - D_{\mu\nu}D_{\mu\nu}- 2 \Lambda^{\alpha \z}\sigma_{m\z \dot \z} \bigl[\phi_m,\Lambda_\alpha^{\dot \z}\bigr] - 2i\Lambda^{\alpha \z}\nabla_{\! w} \Lambda_{\alpha \z}
	\\
	 +2i \Lambda^\alpha_{\dot \z}  \nabla_{\! \bar w} \Lambda_\alpha^{\dot \z} + 2 \bar \Lambda_{\dot \alpha}^\z \sigma_{m \z\dot \z} \bigl[ \phi_m, \bar \Lambda^{\dot \alpha \dot \z} \bigr] -2i \bar \Lambda_{\dot \alpha}^\z \nabla_{\! \bar w} \bar \Lambda^{\dot \alpha}_\z + 2i \bar \Lambda_{\dot \alpha \dot \z} \nabla_{\! w} \bar \Lambda^{\dot \alpha \dot \z}
	\\
	 - 2i \sigma_{\mu \alpha \dot \alpha} \Lambda^\alpha_\z \bigl[ a_\mu, \bar \Lambda^{\dot \alpha \z}\bigr] + 2i \sigma_{\mu \alpha \dot \alpha} \Lambda^{\alpha \dot \z} \bigl[ a_\mu, \bar \Lambda^{\dot \alpha}_{\dot \z}\bigr]
	\biggr)
	\\
	+\frac 12 \nabla_{\! I} \tilde q^{\alpha f}\nabla_{\! I} q_{\alpha f} -\frac 12 \phi_m \tilde q^{\alpha f}\phi_m q_{\alpha f} - \frac 12 \tilde \chi^{f\z} \sigma_{m \z \dot \z} \phi_m \chi^{\dot \z}_f + \frac 12 \tilde \chi^f_{\dot \z} \bar \sigma_m^{\dot \z \z} \phi_m \chi_{f\z}
	\\
	+i \tilde \chi^{f\z} \nabla_{\! \bar w} \chi_{f\z} - i \tilde \chi^f_{\dot \z} \nabla_{\! w} \chi^{\dot \z}_f - \frac 1{\sqrt 2} \tilde q^{\alpha f}\Lambda_{\alpha \z}\chi^\z_f + \frac 1{\sqrt 2} \tilde q^{\alpha f}\Lambda_\alpha^{\dot \z} \chi_{f\dot \z}
	\\
	-\frac 1{\sqrt 2} \tilde \chi^{\z f} \Lambda^\alpha_\z q_{\alpha f} + \frac 1{\sqrt 2} \tilde \chi^f_{\dot \z} \Lambda^{\alpha \dot \z} q_{\alpha f} + \frac i2 \tilde q^{\alpha f} D_{\mu\nu} \sigma_{\mu\nu\alpha}^{\phantom{\mu\nu\alpha} \beta} q_{\beta f}\, .
\end{multline}
We have introduced the usual complex worldsheet coordinates $w$ and $\bar w$, with associated covariant derivatives $\nabla_{\! w}=(\nabla_1-i\nabla_2)/2$ and $\nabla_{\! \bar w}=(\nabla_1+i\nabla_2)/2$. We have also introduced a self-dual auxiliary field $D_{\mu\nu}$ in the adjoint of $\text{U}(K)$. This field allows us to write the quartic interactions between the scalars $(a,q,\tilde q)$ in a simple way which will be useful for performing the scaling limit in the next subsection. The overall normalization as well as the constant term in the Lagrangian are fixed by the D-string tension, the string length being related to the fundamental string tension by $\ls^{2}=2\pi\alpha'$.

\subsection{The scaling limit}

The scalar fields $a_{\mu}$ and $\phi_{m}$ in the worldsheet Lagrangian \eqref{d1lag} are associated with the motion of the D-strings parallel and transverse to the D5-branes respectively. The corresponding coordinates are
\be\label{coorddef} X_{\mu} =  \ls^{2}a_{\mu}\, ,\quad
Y_{m} = \ls^{2}\phi_{m}\, .\ee
To implement the standard decoupling, near-horizon limit \cite{Maldacena:1997re,Itzhaki:1998dd, WittenCFTHiggs}, we take $\ls\rightarrow 0$ while keeping $X_{\mu}$, $Y_{m}/\ls^{2}=\phi_{m}$ and the six-dimensional 't Hooft coupling $N \ls^{2} \gs$ fixed, as usual. Supersymmetry then dictates that the fermionic superpartners $\Lambda_\alpha$ and 
$\bar \psi^{\dot \alpha} = \ls^2 \bar \Lambda^{\dot \alpha}$ of $\phi_{m}$ and $X_{\mu}$ respectively must also be kept fixed. Introducing $\psi_{\alpha} = \ls^{2}\Lambda_{\alpha}$, the Lagrangian \eqref{d1lag} then simplifies in the scaling to
\begin{align}\notag
	L_{\rm p} &= \frac{1}{2\ls^2 \gs}\truK \biggr( 2\1_K+ \nabla_{\! I} X_\mu \nabla_{\! I} X_\nu + 2i \bigl[X_\mu,X_\nu \bigr]D_{\mu\nu} - \ls^{-4}\bigl[Y_m,X_\mu \bigr]\bigl[Y_m,X_\mu \bigr] 
	\\\notag
	&\hspace{3.4cm} -2i \bar  \psi_{\dot \alpha}^\z \nabla_{\! \bar w} \bar  \psi^{\dot \alpha}_\z + 2i \bar  \psi_{\dot \alpha \dot \z}\nabla_{\! w} \bar  \psi^{\dot \alpha \dot \z} + 2\ls^{-2} \bar  \psi_{\dot \alpha}^\z \sigma_{m \z\dot \z} [ Y_m, \bar  \psi^{\dot \alpha \dot \z} ] 
	\\\notag
	&\hspace{4.6cm}- 2i\ls^{-2} \psi^\alpha_\z \sigma_{\mu \alpha \dot \alpha} \bigl[ X_\mu, \bar  \psi^{\dot \alpha \z}\bigr] + 2i\ls^{-2} \psi^{\alpha \dot \z} \sigma_{\mu \alpha \dot \alpha} \bigl[ X_\mu, \bar  \psi^{\dot \alpha}_{\dot \z}\bigr]
	\biggl)
		\\\notag
	&+\frac 12 \nabla_{\! I} \tilde q^{\alpha f}\nabla_{\! I} q_{\alpha f} -\frac {1}{2\ls^{4}} Y_m \tilde q^{\alpha f}Y_m q_{\alpha f} - \frac 1{2\ls^2} \tilde \chi^{f\z} \sigma_{m \z \dot \z} Y_m \chi^{\dot \z}_f + \frac 1{2\ls^2} \tilde \chi^f_{\dot \z} \bar \sigma_m^{\dot \z \z} Y_m \chi_{f\z}
	\\\notag
	&\hspace{3cm}+i \tilde \chi^{f\z} \nabla_{\! \bar w} \chi_{f\z} - i \tilde \chi^f_{\dot \z} \nabla_{\! w} \chi^{\dot \z}_f - \frac 1{\sqrt 2 \ls^2} \tilde q^{\alpha f}\psi_{\alpha \z}\chi^\z_f + \frac 1{\sqrt 2 \ls^2} \tilde q^{\alpha f}\psi_\alpha^{\dot \z} \chi_{f\dot \z}
	\\\label{d1lag1}
	&\hspace{3.8cm}-\frac 1{\sqrt 2 \ls^2} \tilde \chi^{\z f} \psi^\alpha_\z q_{\alpha f} + \frac 1{\sqrt 2 \ls^2} \tilde \chi^f_{\dot \z} \psi^{\alpha \dot \z} q_{\alpha f} + \frac i2 \tilde q^{\alpha f} D_{\mu\nu} \sigma_{\mu\nu\alpha}^{\phantom{\mu\nu\alpha} \beta} q_{\beta f}\, .
\end{align}
This Lagrangian will be our starting point for the probe D-string worldsheet theory.

It is important to emphasize that the scalar fields $Y_m$ are non-dynamical auxiliary variables in \eqref{d1lag1}. Indeed, they do not have a kinetic term and could be trivially integrated out. However, as explained in the next Section, these fields become dynamical due to the quantum corrections and  play a central r\^ole both in the mathematics and the physical interpretation of the solution of the model at large $N$ \cite{Ferrari:2012nw}.

\section{Solving the model at large $N$}
\label{solution}

We now solve the model defined by the Lagrangian \eqref{d1lag1} along the lines of \cite{Ferrari:2012nw,Ferrari:2013pq,Ferrari:2013hg}. The crucial property is that the fields $(q,\tilde q,\chi,\tilde\chi)$ carry only one $\text{U}(N)$ index and are thus vector-like variables. The large $N$ path integral over these fields can then always be performed exactly, using standard techniques for large $N$ vector models \cite{Coleman:1980nk,ZinnJustin:1998cp,Ferrari:2000wq,Ferrari:2001jt,Ferrari:2002gy}. The idea is to rewrite the action by introducing suitable auxiliary fields, in order to make the vector variables appear only quadratically. In our case, the relevant auxiliary fields are precisely the variables $(Y_{m},\psi_{\alpha\zeta},\psi_{\alpha\dot\zeta},D_{\mu\nu})$ which we have already included when writing \eqref{d1lag1}. The path integral over the vector variables is then Gaussian. The result is an effective action for the auxiliary fields, which become dynamical through the quantum loops of the vector variables. Moreover, and most importantly, this effective action is automatically proportional to $N$ because the vector fields have $N$ components. At large $N$, it can thus be treated classically. 

The resulting structure is thus perfectly consistent with the D-string seen as moving in a higher dimensional classical non-trivial background. Indeed, the fields $Y_{m}$ can be interpreted as the emerging coordinates which behave classically at large $N$ and the metric on the emerging space will be related to the kinetic term for the $Y_{m}$.

Let us now carry out this procedure explicitly for our model, mainly focusing on the case of a single D-string probe.

\subsection{Integrating out}
\label{intout}

The effective action $N \Seff$ is given by
\be
\label{defSeff}
	e^{-N\Seff} = \int \d q \d \tilde q \d \chi \d \tilde \chi \ e^{-S_{\rm p}}\, ,
\ee
where $S_{\rm p}$ is the action for the Lagrangian \eqref{d1lag1}. In order to derive the emergent geometry, we can focus on the bosonic part of the effective action and thus set the fermionic fields $\psi$ and $\bar \psi$ to zero. Note, however, that computing the fermionic terms in the effective action could also be done straightforwardly.

The integral \eqref{defSeff} is Gaussian and yields
\begin{multline}
\Seff (A,X,Y,D) =\frac{1}{2N \ls^{2} \gs} \int\! \d^2 w \, \truK\Bigr( 2\1_K + \nabla_{\! I} X_\mu \nabla_{\! I} X_\mu + 2i \bigl[X_\mu,X_\nu\bigr] D_{\mu\nu} 
	\\\label{seff2}- \ls^{-4}\bigl[Y_m,X_\mu\bigr]\bigl[Y_m,X_\mu\bigr] 	\Bigl) 
	+ \ln \Delta_{q,\tilde q} - \ln \Delta_{\chi,\tilde \chi}\, ,
\end{multline}
where the determinants $\Delta_{q,\tilde q}$ and $\Delta_{\chi,\tilde \chi}$ are given by
\begin{align}\label{det1}
	\Delta_{q,\tilde q} &= \det \bigl( -\1_{K}\otimes\1_{2}\, \nabla^2 + \ls^{-4}Y_{m} Y_{m} \otimes \1_2 + i D_{\mu\nu} \otimes \sigma_{\mu\nu} \bigr) \, ,\\\label{det2}
	\Delta_{\chi,\tilde \chi} &= \det
		\begin{pmatrix}
			-2i\1_{K}\otimes\1_{2}\, \nabla_{\!\bar w} & \ls^{-2}Y_m \otimes \sigma_m \\
			- \ls^{-2}Y_m \otimes \bar \sigma_m & 2i\1_{K}\otimes\1_{2}\,\nabla_{\! w}
		\end{pmatrix}\, .
\end{align}
At large $N$, the field $D_{\mu\nu}$ is fixed in terms of the other variables by the saddle point equation
\be\label{saddle} \frac{\delta\Seff}{\delta D_{\mu\nu}} = 0\, .\ee
If we specialize to the case $K=1$ of a single D-string probe, then the solution is simply $D_{\mu\nu}=0$. This follows from the vanishing of the linear term in $D$ in the expansion of \eqref{det1} around $D=0$ or, equivalently, from the commuting nature of the $X_{\mu}$ and $\text{SO}(4)$ invariance. We thus get
\begin{multline}\label{seff3}
	\Seff (A,X,Y) = \frac{1}{N \ls^{2} \gs} \int \d ^2 z \ \Bigr( 1+\frac12 \del_I X_\mu \del_I X_\mu \Bigl) 
	\\ +2\ln \det \bigl( -\nabla^2 + \ls^{-4}Y_{m} Y_{m} \bigr)
	-\ln\det
		\begin{pmatrix}
			-2i\1_{2} \nabla_{\! \bar w} & \ls^{-2}Y_m \sigma_m \\
			- \ls^{-2}Y_m \bar \sigma_m & 2i\1_{2} \nabla_{\! w}
		\end{pmatrix}\, .
\end{multline}
\subsection{The effective action up to cubic order}
\label{micr1}

We are going to use \eqref{seff3} up to order three in an expansion in the field strength $F_{IJ}$ and around constant values of the coordinate worldsheet fields,
\be
\label{exp0}
	X_\mu = x_\mu + \ls^2 \epsilon_\mu\, ,\quad Y_m = y_m + \ls^{2}\epsilon_m \, .\ee
Eventually, we shall match $N\Seff$ up to this order in the next Section with the D-string action in a general type IIB background. 

The explicit computation of the expansion is straightforward. Let us introduce the radial coordinate $r$ defined by
\be\label{radialdef} r^{2} = y_{m}y_{m}\, .\ee
We write
\begin{align}\label{lndet1}
	\ln \Delta_{q,\tilde q} & = 2 \ln \det K_B + 2 \tr \ln (\1+K_{B}^{-1}\varphi)\, ,
	\\\label{lndet2}
	\ln \Delta_{\chi,\tilde \chi} &=  \ln \det K_F + \tr \ln (\1+K_{F}^{-1}\xi)\, ,
\end{align}
in terms of the bosonic and fermionic propagators
\begin{align}\label{kbinv}
	K_B^{-1} (w,w') &= \int\! \frac{\d^2 p}{(2\pi)^2}\frac{e^{ip\cdot (w-w')}}{p^{2}+\ls^{-4}r^{2}}\, \cvp\\\label{kfinv}
	K_F^{-1} (w,w') &= \int\! \frac{\d^2 p}{(2\pi)^2} 
	\frac{e^{ip\cdot (w-w')}}{p^{2}+\ls^{-4}r^{2}}
	\begin{pmatrix}
		2\1_2 p_w & \ls^{-2}y_m \sigma_m \\
		-\ls^{-2}y_m \bar \sigma_m & -2\1_2 p_{\bar w}
	\end{pmatrix}
		\, 
\end{align}
and with
\begin{align}\label{var1}
	\varphi & =  2 \ls^{-2} y_m \epsilon_m + \epsilon_m \epsilon_m-i \del_{I}  A_{I} -2 i A_{I} \partial_{I} + A_{I}A_{I}\, ,
	\\ \label{var2}
	\xi & = 
	\begin{pmatrix}
		2\1_2 A_{\bar w} & \epsilon_m \sigma_m \\
		-\epsilon_m \bar \sigma_m & -2\1_2 A_w
	\end{pmatrix}\, .
\end{align}
We then expand the traces in \eqref{lndet1} and \eqref{lndet2} using
\be\label{lnexp}
	\ln (\1+\delta) = -\sum_{k=1}^\infty \frac{(-1)^{k}}{k} \delta^k
\ee
and compute the resulting one-loop Feynman integrals.

At zeroth and first order, the contributions from the bosonic and fermionic determinants cancel each other and only the constant D-string tension term in \eqref{seff3} remains. At second and third order, we obtain a non-local effective action, which we write as a power series in derivatives by expanding the associated Feynman integrals for small external momenta. Overall, we get
\begin{multline}\label{Seffsol} N\Seff (A,X,Y) = \int\!\d^{2}w\,
\Bigl( \frac{1}{ \ls^{2}\gs} + \frac{\ls^2}{2 \gs } \del_I \epsilon_\mu\del_I \epsilon_\mu +\frac{N\ls^{4}}{4\pi r^2} \del_I \epsilon_m\del_I \epsilon_m + \frac{N\ls^{4}}{8\pi r^2} F_{IJ}F_{IJ}\\
-\frac{N\ls^{6}}{2\pi r^4}y_{m} \epsilon_m \del_I \epsilon_n\del_I \epsilon_n  
	- \frac{N\ls^{6}}{4\pi r^4} \epsilon_m y_{m} F_{IJ}F_{IJ}
	- \frac{iN\ls^{6}}{6\pi r^4} \epsilon_{IJ} y_{m} \epsilon_{mnlp} \epsilon_n \del_I \epsilon_l \del_J \epsilon_p\Bigr) + \cdots 
\end{multline}
where the $\cdots$ stand for terms of quartic or higher order and terms with  more than two derivatives. Of course, the result is consistent with the symmetries of the microscopic theory discussed in Section \ref{sixtotwo}, including the worldsheet parity which must come accompanied by a parity transformation in the directions $y_{m}$.

\section{The emergent background}
\label{comparison}

The action for a D-string moving in a general type IIB supergravity background is the sum of a Dirac-Born-Infeld term and a Chern-Simons term \cite{Leigh:1989jq,Douglas:1995bn}, 
\be\label{dsact1}
S=\frac{1}{\ls^2\gs}\int\! \d^2 \xi\, e^{-\Phi} \sqrt{\det\left[\text{P}(G+B)+\ls^2 F \right]}+\frac{i}{\ls^2\gs}\int \Bigl[\text{P}(C_0 B + C_{2}) +\ls^2 C_0 F \Bigr]
\, ,\ee
where the fields $\Phi$, $G$, $B$, $C_{0}$ and $C_{2}$ are the dilaton, string-frame metric, Kalb-Ramond two-form and Ramond-Ramond potentials respectively, $F$ is the worldsheet field strength and P denotes the pull-back of the spacetime fields to the worldsheet. Working in the static gauge and writing the fields $Z_{i}$, $1\leq i\leq 8$, corresponding to the coordinates transverse to the D-string worldsheet as
\be\label{zexp} Z_{i} = z_{i}  + \ls^{2}\epsilon_{i}\, ,\ee
we can expand \eqref{dsact1} in powers of $\epsilon_{i}$ and $F$. Following the basic idea that underlies the present work as well as our previous studies \cite{Ferrari:2012nw, Ferrari:2013pq,Ferrari:2013hg}, this expansion should match with the similar expansion \eqref{Seffsol} of the effective action  describing the solution of the large $N$ microscopic model of the D-strings in the presence of the $N$ D5-branes. Morevover, we should be able to derive the precise supergravity background sourced by the D5-branes from the coefficients in the expansion \eqref{Seffsol}. Let us check that this is indeed the case.

The zeroth order Lagrangian derived in this way from \eqref{dsact1} reads
\be\label{o0}
	L^{(0)} = \frac{1}{\ls^2 \gs} 
		\Bigl[
			e^{-\Phi} \sqrt{\det (G_{IJ}+B_{IJ})} 
			+
			\frac{i}{2} \epsilon^{IJ}\bigl( C_0 B_{IJ} + (C_2)_{IJ} \bigr)
		\Bigr]\, ,
\ee
where the capital Latin indices $1\leq I,J,\ldots\leq 2$ correspond as usual to the directions parallel to the D-string worldsheet. Matching with the microscopic result \eqref{Seffsol} and taking into account worldsheet parity invariance yields the conditions
\be\label{pred00prout}
	e^{-\Phi}\sqrt{\det(G_{IJ}+B_{IJ})}=1\, ,\quad
 C_0 B_{IJ} + (C_2)_{IJ} = 0\, .\ee
Taking these constraints into account, the Lagrangian derived from \eqref{dsact1} at first order in $\epsilon$ is, up to an irrelevant total derivative,
\be\label{o1}
	L^{(1)} = \frac{1}{2 \gs} F_{IJ}\bigl( -E^{IJ} + iC_0 \epsilon^{IJ} \bigr)\,,
\ee
where the matrix $E^{IJ}$ is the inverse of $G_{IJ}+B_{IJ}$. Using the fact that $L^{(1)}=0$ in \eqref{Seffsol} and worldsheet parity, we get
\be\label{pred10pffuit} B_{IJ}= 0\, ,\quad  C_0 = 0 \, .\ee
The conditions \eqref{pred00prout} thus reduce to
\be\label{pred01}
	e^{-\Phi}\sqrt{\det G_{IJ}}=1\, , \quad (C_2)_{IJ} = 0\, .
\ee
Taking these results into account as well as the fact that $G_{iI}=0$ from $\text{ISO}(2)$ invariance along the worldsheet, the second order Lagrangian derived from \eqref{dsact1} then reads
\begin{multline}\label{o2}
L^{(2)} = \frac{\ls^2}{\gs}
		\biggl[\bigl( \frac 12 \tilde G^{IJ}G_{ij} + \tilde G^{I[J}\tilde G^{K]L}B_{iL}B_{jK}\bigr)\,\del_I \e_i \del_J \e_j
			\\+ \frac{1}{4}\tilde G^{IJ}\tilde G^{KL}\, F_{IK}F_{JL}
			-i \epsilon^{IJ} \del_{[i} (C_2)_{j]I}\,\e_{i}\del_J \e_{j}
		\biggr]\, ,
\end{multline}
where the matrix $\tilde G^{IJ}$ is the inverse of the two-by-two matrix $G_{IJ}$. Comparing with \eqref{Seffsol} and using again parity invariance yields
\be
\label{praat}G_{IJ} = \sqrt{\frac{2\pi r^2}{\ls^2 \gs N}}\,\delta_{IJ}\, , \quad \del_{[i}(C_2)_{j] I} = 0\ee
and then
\be\label{dilaton}
e^{\Phi} = \sqrt{\frac{2\pi r^2}{\ls^2 \gs N}} 
\ee
by using \eqref{pred01}. Comparing the terms $\del_I \e_i \del_J \e_j$
in \eqref{Seffsol} and \eqref{o2}, using \eqref{praat} and the $\text{ISO}(2)$ symmetry to fix $B_{Ii}=0$, we find
\be\label{Gmunumn}
G_{\mu\nu} = \sqrt{\frac{2\pi r^2}{\ls^2 \gs N}}\delta_{\mu\nu}\, ,\quad
G_{mn} = \sqrt{\frac{\ls^2 \gs N}{2 \pi  r^2}}\delta_{mn}\, .
\ee
We thus find that the components $G_{IJ}$ and $G_{\mu\nu}$ match, which shows that the metric has the expected $\text{SO}(6)$ isometry of the background sourced by D5-branes.
We can continue the same analysis at third order. Using the constraints on the background that we have already derived, \eqref{dsact1} yields the third order Lagrangian up to two derivative terms,
\begin{multline}\label{o3}
	L^{(3)} = \frac{\ls^4}{2\gs}
		\biggl[ \del_i \bigl(\tilde G^{IJ}G_{jk}\bigr)\,
				\e_i\del_I \e_j \del_J \e_k  
	+i\epsilon^{IJ}\del_{[i} (C_2)_{jk]}\,\e_i \del_I \e_j \del_J \e_k 
			\\
			+\frac{1}{2}  \del_i (\tilde G^{IJ}\tilde G^{KL}) \, \e_i F_{IL}F_{JK} 
			+\tilde G^{IK}\tilde G^{JL} B_{ij}\,  F_{KL}\del_I \e_i \del_J \e_j	\biggr]\, .
\end{multline}
Matching with \eqref{Seffsol} and using the $\text{ISO}(2)$ invariance and $\text{SO}(6)$ isometry of the background, we then obtain 
\be
\label{Bijsol}
	B_{ij} = 0\, ,\quad F_3 = \d C_2 = \frac 16 \epsilon_{mnlp}\del_p e^{-2\Phi} \d y_m \wedge \d y_n \wedge \d y_l\, .
\ee
Overall, we have derived the following type IIB supergravity background
\begin{align}\notag
	e^{\Phi} &= \sqrt{\frac{2 \pi r^2}{\ls^2 g_s N}}\, \cvp \quad \d s^2 = e^{\Phi}  (\d w_I\d w_I + \d x_\mu\d x_\mu) + e^{-\Phi} \d y_m \d y_m\, , \\\label{summary}
	F_3 &= \d C_2 = \frac 16 \epsilon_{mnlp}\del_p e^{-2\Phi} \d y_m \wedge \d y_n \wedge \d y_l\,, \quad B = C_0 = 0\,,
\end{align}
which perfectly matches with the well-known near-horizon geometry of $N$ D5-branes \cite{Horowitz:1991cd}.

\subsection*{Acknowledgements}
We would like to thank Micha Moskovic for useful discussions.
This work is supported in part by the Belgian Fonds de la Recherche
Fondamentale Collective (grant 2.4655.07) and the Belgian Institut
Interuniversitaire des Sciences Nucl\'eaires (grant 4.4511.06 and 4.4514.08). A.R. is a Research Fellow of the Belgian Fonds de la Recherche Scientifique-FNRS.
\appendix
\section{Notations and conventions}
\label{notconv}
We work in Euclidean signature throughout this paper. All our conventions are chosen consistently with those of \cite{Ferrari:2012nw,Ferrari:2013pq,Ferrari:2013hg}.

\begin{table}
\be\nonumber
\begin{matrix}
& \text{Spin}(4) & \text{Spin}(4)' &\text{U}(1) & \text{U}(N) & \text{U}(K)\\
\hline
I,J,\ldots & (0,0) & (0,0) & 1 & \mathbf 1 & \mathbf 1\\
\mu, \nu, \ldots & (1/2,1/2) & (0,0) & 0 & \mathbf 1&\mathbf 1 \\
m, n, \ldots & (0,0) & (1/2,1/2) & 0 & \mathbf 1&\mathbf 1 \\
\alpha, \beta, \ldots \ \text{(upper or lower)}& (1/2,0) & (0,0) & 0 & \mathbf 1&\mathbf 1 \\
\dot\alpha, \dot\beta, \ldots \ \text{(upper or lower)}& (0,1/2) & (0,0) & 0 & \mathbf 1&\mathbf 1 \\
\z, \xi, \ldots \ \text{(upper or lower)}& (0,0) & (1/2,0) & 0 & \mathbf 1&\mathbf 1 \\
\dot \z, \dot \xi, \ldots \ \text{(upper or lower)}& (0,0) & (0,1/2) & 0 & \mathbf 1&\mathbf 1 \\
f, f', \ldots \ \text{(lower)}& (0,0) & (0,0) & 0 & \mathbf N&\mathbf 1 \\
f, f', \ldots \ \text{(upper)}& (0,0) & (0,0) & 0 & \mathbf{\bar N}&\mathbf 1 \\
i, j, \ldots \ \text{(lower)}& (0,0) & (0,0) & 0 & \mathbf 1&\mathbf K \\
i, j, \ldots \ \text{(upper)}& (0,0) & (0,0) & 0 & \mathbf 1&\mathbf{\bar K}\\
A_I & (0,0) & (0,0) & 1 & \mathbf 1 & \textbf{Adj} \\
X_{\mu i}^{\phantom{\mu i}j}=\ls^{2}A_{\mu i}^{\phantom{\mu i}j} & (1/2,1/2) & (0,0) & 0 & \mathbf 1 & \textbf{Adj}\\
Y_{m i}^{\phantom{\mu i}j}=\ls^{2}\phi_{m i}^{\phantom{\mu i}j} & (0,0) & (1/2,1/2) & 0 & \mathbf 1& \textbf{Adj} \\
\psi_{\alpha \z i}^{\phantom{\alpha \z i} j} =\ls^{2}\Lambda_{\alpha \z i}^{\phantom{\alpha \z i} j} & (1/2,0) & (1/2,0) & 1/2 & \mathbf 1  & \textbf{Adj}\\
\psi^{\phantom{\alpha \dot \z i} j}_{\alpha \dot \z i} =\ls^{2}\Lambda^{\phantom{\alpha \dot \z i} j}_{\alpha \dot \z i} & (1/2,0) & (0,1/2) & \mathllap{-}1/2 & \mathbf 1  & \textbf{Adj}\\
\bar\psi^{\dot\alpha \phantom{\z i}j}_{\phantom{\dot\alpha}\z i} =\ls^{2}\bar\Lambda^{\dot\alpha \phantom{\z i}j}_{\phantom{\dot\alpha}\z i} & (0,1/2) & (1/2,0) & \mathllap{-}1/2 & \mathbf 1 & \textbf{Adj} \\
\bar\psi^{\dot\alpha \phantom{\dot \z i} j}_{\phantom{\dot\alpha} \dot \z i} =\ls^{2}\bar\Lambda^{\dot\alpha \phantom{\dot \z i} j}_{\phantom{\dot\alpha} \dot \z i} & (0,1/2) & (0,1/2) & 1/2 & \mathbf 1 & \textbf{Adj} \\
D_{\mu\nu i}^{\ \ \ j} & (1,0) & (0,0) & 0 & \mathbf 1 & \textbf{Adj}\\
q_{\alpha f i} & (1/2,0) & (0,0) & 0 & \mathbf N  & \textbf{K}\\
\tilde q^{\alpha f i} & (1/2,0) & (0,0) & 0 & \mathbf{\bar N}  & \mathbf{\bar K}\\
\chi_{\z f i} & (0,0) & (1/2,0) & \mathllap{-}1/2 & \mathbf{N}  & \mathbf{K}\\
\chi_{\dot \z f i} & (0,0) & (0,1/2) & 1/2 & \mathbf{N}  & \mathbf{K}\\
\tilde\chi^{\phantom{\z} fi}_\z & (0,0) & (1/2,0) & \mathllap{-}1/2 & \mathbf{\bar N}  & \mathbf{\bar K}\\
\tilde\chi_{\dot \z}^{fi} & (0,0) & (0,1/2) & 1/2 & \mathbf{\bar N}  & \mathbf{\bar K}
\end{matrix}
\ee
\caption{\label{recapsym}Conventions for the transformation laws of indices and fields. For maximum clarity, we have indicated all the indices associated to each field, whereas in the main text the gauge $\text{U}(N)$ and $\text{U}(K)$ indices are usually suppressed. The representations of $\text{Spin}(4)=\text{SU}(2)_{+}\times\text{SU}(2)_{-}$ and $\text{Spin}(4)'=\text{SU}(2)'_{+}\times\text{SU}(2)'_{-}$ are indicated according to the spin in each $\text{SU}(2)$ factor. The $(1/2,1/2)$ of $\text{SU}(2)_{+}\times\text{SU}(2)_{-}$ corresponds to the fundamental representations of $\text{SO}(4)$. The $\text{U}(1)$ group corresponds to the worldsheet rotations under which positive and negative chirality spinors have charge $1/2$ and $-1/2$ respectively.}
\end{table}
\subsection{Symmetries and indices}
For $K$~D-strings parallel to $N$~D$5$-branes, the symmetry group is $\SO(2) \times \SO(4) \times \SO(4)'$. The $\SO(2)$ factor corresponds to rotations in the two-dimensional space parallel to the D-strings, the $\SO(4)$ factor corresponds to rotations in the four-dimensional space parallel to the D5-branes and transverse to the D-strings and the $\SO(4)'$ factor corresponds to rotations in the four-dimensional space transverse to both the D-strings and the D5-branes.

We use indices $1\leq I,J,\ldots\leq 2$ for the vectors of $\SO(2)$, $1\leq \mu,\nu \leq 4$ and $1\leq \alpha,\beta \leq 2$ for the vectors and left-handed spinors of $\SO(4)$ respectively and $1\leq m,n \leq 4$ and $1\leq \z,\xi \leq 2$ for the vectors and left-handed spinors of $\SO(4)'$ respectively. For right-handed spinors we use dotted indices $\dot\alpha$, $\dot\z$ etc. The SO(4) spinor indices are raised and lowered according to the standard conventions
\begin{align}
	\lambda_\alpha &= \epsilon_{\alpha \beta} \lambda^{\beta} \,,\\
	\psi^{\dot \alpha} &= \epsilon^{\dot \alpha \dot \beta} \psi_{\dot \beta} \,,
\end{align}
with $\epsilon_{21}=-\epsilon_{12}=-\epsilon^{21}=\epsilon^{12}=1$, and similarly for the SO(4)$'$ spinor indices $\z,\xi,$ etc.

We also have gauge group indices $f,f'$ and $i,j$ for $\UN$ and $\UK$ respectively. We often suppress these indices if there is no ambiguity.

The branes are located in $\mathbb R^{10}$ according to Table \ref{loc}. We denote $w_I$ the two coordinates parallel to the D-strings and by $(z^i)=(x_\mu,y_m)$ the coordinates transverse to the D-strings. We also define the radial coordinate $r$ by
\be\label{rdef}
	r^2 = y_m y_m \, .
\ee
\begin{table}[h]
\be
\begin{array}{|c|c c c c c c c c c c|}
	\hline
     & 1 & 2 & 3 & 4 & 5 & 6 & 7 & 8 & 9 & 10\\
    \hline
    {\rm D}5 & \times & \times & \times & \times & \times & \times & & & & \\
    \hline
    {\rm D}1 & \times & \times & & & & & & & & \\
    \hline
     & w_1 & w_2 & x_1 & x_2 & x_3 & x_4 & y_1 & y_2 & y_3 & y_4\\
    \hline
\end{array}
\ee\caption{\label{loc}Location of the D1- and D5-branes in $\mathbb R^{10}$. The third row indicates the notation we use for the various types of coordinates.}
\end{table}
\subsection{Four-dimensional algebra}

The standard Pauli matrices are taken to be
\begin{equation}
  \sigma_1=\begin{pmatrix}0 & 1\\1 & 0\end{pmatrix},\quad \sigma_2=\begin{pmatrix}0 & -i\\i & 0\end{pmatrix}, \quad \sigma_3=\begin{pmatrix}1 & 0\\0 & -1\end{pmatrix}
  \label{paulidef}\, .
\end{equation}
We define
\begin{equation}
  \sigma_{\mu\alpha\dot\alpha}=(\vec\sigma,-i\1_{2\times2})_{\alpha\dot\alpha}\,, \quad \bar\sigma_\mu^{\dot\alpha\alpha}=(-\vec\sigma,-i\1_{2\times2})^{\dot\alpha\alpha}
  \label{sigma1def}
\end{equation}
and
\begin{equation}
  \sigma_{\mu\nu}=\frac14(\sigma_{\mu}\bar\sigma_{\nu}-\sigma_{\nu}\bar\sigma_{\mu})\,, \quad 
  \bar\sigma_{\mu\nu}=\frac14(\bar\sigma_{\mu}\sigma_{\nu}-\bar\sigma_{\nu}\sigma_{\mu})\, .
  \label{sigma2def}
\end{equation}
We have the useful following identities for the computation of the effective action at second and third order:
\begin{align}
	\tr (\sigma_\mu \bar \sigma_\nu \sigma_\rho \bar \sigma_\lambda) &=  2(\delta_{\mu \nu}\delta_{\rho \lambda} - \delta_{\mu \rho}\delta_{\nu \lambda} + \delta_{\mu \lambda}\delta_{\rho \nu} - \epsilon_{\mu\nu\rho\lambda}) 
	\, ,	\\
\notag 	a_{\mu_1}a_{\mu_2}a_{\mu_3}\tr(\sigma_{\mu_1}\bar \sigma_{\nu_1} \sigma_{\mu_2} \bar \sigma_{\nu_2} \sigma_{\mu_3}\bar \sigma_{\nu_3} &+ \bar\sigma_{\mu_1}\sigma_{\nu_1} \bar\sigma_{\mu_2} \sigma_{\nu_2} \bar\sigma_{\mu_3}\sigma_{\nu_3} ) =
	\\
	4\bigl[ a^2& (a_{\nu_1}\delta_{\nu_2\nu_3}+a_{\nu_2}\delta_{\nu_1\nu_3}+a_{\nu_3}\delta_{\nu_1\nu_2}) - 4 a_{\nu_1}a_{\nu_2}a_{\nu_3} \bigr] \, ,
\end{align}
for any numbers $a_{\mu}$ and where we set $a^2 = a_\mu a_\mu$ and $\epsilon_{1234}=+1$.
\bibliographystyle{utphys}
\bibliography{emergentbiblio}

\providecommand{\href}[2]{#2}\begingroup\raggedright\begin{thebibliography}{10}

\bibitem{Maldacena:1997re}
J.~M. Maldacena, ``{The Large $N$ Limit of Superconformal Field Theories and
  Supergravity},'' {\em Adv.\ Theor.\ Math.\ Phys.} {\bfseries 2} (1998)
  231--252,
\href{http://arxiv.org/abs/hep-th/9711200}{{\ttfamily arXiv:hep-th/9711200}}.
%%CITATION = HEP-TH/9711200;%%.

\bibitem{Gubser:1998bc}
S.~Gubser, I.~R. Klebanov, and A.~M. Polyakov, ``{Gauge Theory Correlators from
  Noncritical String Theory},''
  \href{http://dx.doi.org/10.1016/S0370-2693(98)00377-3}{{\em Phys.\ Lett.}
  {\bfseries B428} (1998) 105--114},
\href{http://arxiv.org/abs/hep-th/9802109}{{\ttfamily arXiv:hep-th/9802109}}.
%%CITATION = HEP-TH/9802109;%%.

\bibitem{Witten:1998qj}
E.~Witten, ``{Anti-de~Sitter Space and Holography},'' {\em Adv.\ Theor.\ Math.\
  Phys.} {\bfseries 2} (1998) 253--291,
\href{http://arxiv.org/abs/hep-th/9802150}{{\ttfamily arXiv:hep-th/9802150}}.
%%CITATION = HEP-TH/9802150;%%.

\bibitem{Aharony:1999ti}
O.~Aharony, S.~S. Gubser, J.~M. Maldacena, H.~Ooguri, and Y.~Oz, ``{Large $N$
  Field Theories, String Theory and Gravity},''
  \href{http://dx.doi.org/10.1016/S0370-1573(99)00083-6}{{\em Phys.\ Rept.}
  {\bfseries 323} (2000) 183--386},
\href{http://arxiv.org/abs/hep-th/9905111}{{\ttfamily arXiv:hep-th/9905111}}.
%%CITATION = HEP-TH/9905111;%%.

\bibitem{D'Hoker:2002aw}
E.~D'Hoker and D.~Z. Freedman, ``{Supersymmetric Gauge Theories and the AdS/CFT
  Correspondence},''
\href{http://arxiv.org/abs/hep-th/0201253}{{\ttfamily arXiv:hep-th/0201253}}.
%%CITATION = HEP-TH/0201253;%%.

\bibitem{Ferrari:2012nw}
F.~Ferrari, ``{Emergent Space and the Example of
  $\text{AdS}_5\times\text{S}^5$},''
  \href{http://dx.doi.org/10.1016/j.nuclphysb.2012.12.004}{{\em Nucl.\ Phys.}
  {\bfseries B869} (2013) 31--55},
\href{http://arxiv.org/abs/1207.0886}{{\ttfamily arXiv:1207.0886 [hep-th]}}.
%%CITATION = ARXIV:1207.0886;%%.

\bibitem{Ferrari:2013pq}
F.~Ferrari, M.~Moskovic, and A.~Rovai, ``{Examples of Emergent Type IIB
  Backgrounds from Matrices},''
\href{http://arxiv.org/abs/1301.3738}{{\ttfamily arXiv:1301.3738 [hep-th]}}.
%%CITATION = ARXIV:1301.3738;%%.

\bibitem{Ferrari:2013pi}
F.~Ferrari, ``{On Matrix Geometry and Effective Actions},''
\href{http://arxiv.org/abs/1301.3722}{{\ttfamily arXiv:1301.3722 [hep-th]}}.
%%CITATION = ARXIV:1301.3722;%%.

\bibitem{Ferrari:2013hg}
F.~Ferrari and M.~Moskovic, ``{Emergent D4-Brane Background from
  D-Particles},''
\href{http://arxiv.org/abs/1301.7062}{{\ttfamily arXiv:1301.7062 [hep-th]}}.
%%CITATION = ARXIV:1301.7062;%%.

\bibitem{Itzhaki:1998dd}
N.~Itzhaki, J.~M. Maldacena, J.~Sonnenschein, and S.~Yankielowicz,
  ``{Supergravity and the Large $N$ Limit of Theories with Sixteen
  Supercharges},'' \href{http://dx.doi.org/10.1103/PhysRevD.58.046004}{{\em
  Phys.\ Rev.} {\bfseries D58} (1998) 046004},
\href{http://arxiv.org/abs/hep-th/9802042}{{\ttfamily arXiv:hep-th/9802042
  [hep-th]}}.
%%CITATION = HEP-TH/9802042;%%.

\bibitem{Green:2000ke}
M.~B. Green and M.~Gutperle, ``{D-Instanton Induced Interactions on a
  D3-Brane},'' {\em JHEP} {\bfseries 0002} (2000) 014,
\href{http://arxiv.org/abs/hep-th/0002011}{{\ttfamily arXiv:hep-th/0002011}}.
%%CITATION = HEP-TH/0002011;%%.

\bibitem{Billo:2002hm}
M.~Bill\'o, M.~Frau, F.~Fucito, A.~Lerda, A.~Liccardo, and I.~Pesando,
  ``{Classical Gauge Instantons from Open Strings},'' {\em JHEP} {\bfseries
  0302} (2003) 045,
\href{http://arxiv.org/abs/hep-th/0211250}{{\ttfamily arXiv:hep-th/0211250}}.
%%CITATION = HEP-TH/0211250;%%.

\bibitem{Diaconescu:1997ut}
D.-E. Diaconescu and R.~Entin, ``{A Nonrenormalization theorem for the d = 1,
  N=8 vector multiplet},''
  \href{http://dx.doi.org/10.1103/PhysRevD.56.8045}{{\em Phys.\ Rev.}
  {\bfseries D56} (1997) 8045--8052},
\href{http://arxiv.org/abs/hep-th/9706059}{{\ttfamily arXiv:hep-th/9706059
  [hep-th]}}.
%%CITATION = HEP-TH/9706059;%%.

\bibitem{frankonematrix}
F.~Ferrari, ``{Emergent Space in Matrix Theories},'' {\em to appear} .

\bibitem{WittenCFTHiggs}
E.~Witten, ``{On the conformal field theory of the Higgs branch},'' {\em JHEP}
  {\bfseries 9707} (1997) 003,
\href{http://arxiv.org/abs/hep-th/9707093}{{\ttfamily arXiv:hep-th/9707093
  [hep-th]}}.
%%CITATION = HEP-TH/9707093;%%.

\bibitem{Coleman:1980nk}
S.~R. Coleman, ``{$1/N$},'' in {\em {Aspects of Symmetry}}.
\newblock {Cambridge U.\ Press},
1985.
\newblock
%%CITATION = SLAC-PUB-2484 ETC.;%%.

\bibitem{ZinnJustin:1998cp}
J.~Zinn-Justin, ``{Vector Models in the Large $N$ Limit: a Few Applications},''
\href{http://arxiv.org/abs/hep-th/9810198}{{\ttfamily arXiv:hep-th/9810198}}.
%%CITATION = HEP-TH/9810198;%%.

\bibitem{Ferrari:2000wq}
F.~Ferrari, ``{Nonsupersymmetric Cousins of Supersymmetric Gauge Theories:
  Quantum Space of Parameters and Double Scaling Limits},''
  \href{http://dx.doi.org/10.1016/S0370-2693(00)01302-2}{{\em Phys.\ Lett.}
  {\bfseries B496} (2000) 212--217},
\href{http://arxiv.org/abs/hep-th/0003142}{{\ttfamily arXiv:hep-th/0003142}}.
%%CITATION = HEP-TH/0003142;%%.

\bibitem{Ferrari:2001jt}
F.~Ferrari, ``{A Model for Gauge Theories with Higgs Fields},'' {\em JHEP}
  {\bfseries 0106} (2001) 057,
\href{http://arxiv.org/abs/hep-th/0102041}{{\ttfamily arXiv:hep-th/0102041}}.
%%CITATION = HEP-TH/0102041;%%.

\bibitem{Ferrari:2002gy}
F.~Ferrari, ``{Large $N$ and Double Scaling Limits in Two-Dimensions},'' {\em
  JHEP} {\bfseries 0205} (2002) 044,
\href{http://arxiv.org/abs/hep-th/0202002}{{\ttfamily arXiv:hep-th/0202002}}.
%%CITATION = HEP-TH/0202002;%%.

\bibitem{Leigh:1989jq}
R.~Leigh, ``{Dirac-Born-Infeld Action from Dirichlet Sigma Model},''
\href{http://dx.doi.org/10.1142/S0217732389003099}{{\em Mod.Phys.Lett.}
  {\bfseries A4} (1989) 2767}.
%%CITATION = MPLAE,A4,2767;%%.

\bibitem{Douglas:1995bn}
M.~R. Douglas, ``{Branes within branes},''
\href{http://arxiv.org/abs/hep-th/9512077}{{\ttfamily arXiv:hep-th/9512077
  [hep-th]}}.
%%CITATION = HEP-TH/9512077;%%.

\bibitem{Horowitz:1991cd}
G.~T. Horowitz and A.~Strominger, ``{Black Strings and P-Branes},''
\href{http://dx.doi.org/10.1016/0550-3213(91)90440-9}{{\em Nucl.\ Phys.}
  {\bfseries B360} (1991) 197--209}.
%%CITATION = NUPHA,B360,197;%%.

\end{thebibliography}\endgroup
\end{document}